\documentclass{ecai} 

\usepackage{latexsym}
\usepackage{amssymb}
\usepackage{amsmath}
\usepackage{amsthm}
\usepackage{booktabs}
\usepackage{enumitem}
\usepackage{graphicx}
\usepackage{color}
\usepackage{multirow}

\newcommand{\BibTeX}{B\kern-.05em{\sc i\kern-.025em b}\kern-.08em\TeX}

\begin{document}

\begin{frontmatter}

\paperid{123} 

\title{Automating AI Failure Tracking: Semantic Association of Reports in AI Incident Database}

\author[A]{\fnms{Diego}~\snm{Russo}\orcid{0009-0007-1095-5168}}
\author[B]{\fnms{Gian Marco}~\snm{Orlando}\orcid{0009-0004-7136-1804}}
\author[C]{\fnms{Valerio}~\snm{La Gatta}\orcid{0000-0002-5941-4684}} 
\author[B]{\fnms{Vincenzo}~\snm{Moscato}\orcid{0000-0002-0754-7696}} 

\address[A]{University of Bergamo, Department of Management, Information and Production Engineering, Via Pasubio 7b, Dalmine (BG), 24044, Italy}
\address[B]{University of Naples Federico II, Department of Electrical Engineering and Information Technology, via Claudio 21, 80125, Naples, Italy}
\address[C]{Northwestern University, Department of Computer Science, McCormick School of Engineering and Applied Science}

\begin{abstract}

Artificial Intelligence (AI) systems are transforming critical sectors such as healthcare, finance, and transportation, enhancing operational efficiency and decision-making processes. However, their deployment in high-stakes domains has exposed vulnerabilities that can result in significant societal harm. To systematically study and mitigate these risk, initiatives like the AI Incident Database (AIID) have emerged, cataloging over 3,000 real-world AI failure reports. Currently, associating a new report with the appropriate \textit{AI Incident} relies on manual expert intervention, limiting scalability and delaying the identification of emerging failure patterns.

To address this limitation, we propose a retrieval-based framework that automates the association of new reports with existing \textit{AI Incidents} through semantic similarity modeling. We formalize the task as a ranking problem, where each report—comprising a title and a full textual description—is compared to previously documented \textit{AI Incidents} based on embedding cosine similarity. Benchmarking traditional lexical methods, cross-encoder architectures, and transformer-based sentence embedding models, we find that the latter consistently achieve superior performance. Our analysis further shows that combining titles and descriptions yields substantial improvements in ranking accuracy compared to using titles alone. Moreover, retrieval performance remains stable across variations in description length, highlighting the robustness of the framework. Finally, we find that retrieval performance consistently improves as the training set expands. Our approach provides a scalable and efficient solution for supporting the maintenance of the AIID.

\end{abstract}

\end{frontmatter}

\section{Introduction}


Artificial Intelligence (AI) systems are increasingly deployed across high-stakes domains, where their decisions significantly impact human lives and societal structures. In healthcare, they have improved diagnostic accuracy and accelerated treatment workflows \cite{chaurasia2022diagnostic, chopra2023revolutionizing, zhao2024haiformer}. Financial institutions leverage AI for credit scoring, fraud detection, and algorithmic trading, achieving gains in efficiency and risk management \cite{addy2024ai, bello2023ai, cohen2022algorithmic, diaconescu2020credit}. AI also underpins key capabilities in autonomous transportation \cite{dartmann2021smart}, criminal justice systems \cite{rigano2019using}, large-scale content moderation \cite{gillespie2020content}, and military operations, where it supports surveillance, target recognition, and mission planning \cite{kase2022future}.

However, real-world deployments have shown that AI systems can fail in unpredictable, opaque, and sometimes harmful ways, especially when deployed at scale. For instance, the use of machine learning to support clinical decision-making has been shown to exacerbate existing health disparities, with certain models underperforming for specific demographic groups \cite{pfohl2021empirical}. Similarly, in the financial sector, AI-driven credit scoring systems have inadvertently resulted in discriminatory outcomes against minority populations \cite{hurley2016credit}. Additional failures span from accidents involving autonomous driving systems\footnote{\url{https://incidentdatabase.ai/cite/23/}} to wrongful arrests caused by facial recognition misidentifications\footnote{\url{https://incidentdatabase.ai/cite/74/}}, emphasizing the potential of AI to reinforce systemic biases, including racism\footnote{\url{https://incidentdatabase.ai/cite/60/}} and misogyny\footnote{\url{https://incidentdatabase.ai/cite/47/}}.

Given the growing evidence of AI failures across diverse application domains, there is a critical need to systematically document and study these incidents to prevent their recurrence and support the development of safer, more reliable systems. Similar to long-established safety infrastructures in aviation and cybersecurity (e.g., FAA, NASA ASRS, CVE), the AI Incident Database (AIID) was established as a structured repository of real-world failures involving AI technologies \cite{McGregor_2021}. By cataloging detailed incident reports, AIID promotes transparency, accountability, and continuous improvement in AI development. Maintained by a coalition of academic, industrial, and non-profit stakeholders, the database currently hosts over 3,000 curated reports, enabling systematic investigation of failure patterns with the goal of reducing the likelihood that similar failures recur. At the core of the AIID is the concept of \textit{AI Incident}, an alleged harm or near-harm event in which an AI system is implicated \cite{mcgregor2022indexingairisksincidents}. Each incident is typically described through one or more reports — narrative accounts derived from journalistic sources, academic literature, or institutional investigations — which provide details about the context, causes, consequences, and actors involved. These reports consist of a title and a free-text description, both of which can vary considerably in length and level of detail, and are complemented by structured metadata such as system functionality, affected stakeholders, and harm types.

However, when a new report is submitted to the AIID, its association with an existing \textit{AI Incident} is still handled manually by human editors \cite{paeth2024lessonseditorsaiincidents}. This reliance on manual expert intervention poses a scalability bottleneck as the volume of incident reports continues to grow, and it introduces potential inconsistencies due to subjective judgment in the classification process. Moreover, the lack of automation limits the ability to rapidly analyze emerging failure patterns across related incidents.

In this work, we address this gap by proposing a method to  assist AIID curators in the task of automatically linking newly submitted reports to existing \textit{AI Incidents}. Specifically, we frame the problem as a retrieval task: the textual content of each report — including its title and description — is compared to previously recorded \textit{AI Incidents} through semantic similarity-based ranking. The goal is to rank historical incidents based on their semantic relevance to the new report. To assess the robustness and generalizability of our approach, we systematically evaluate ranking performance under a variety of conditions — including different input representations (e.g., title only vs. title and description), variations in description length, and temporal dynamics. To operationalize our investigation, we formulate the following research questions (RQs):

\begin{itemize}
    \item[\textbf{RQ1:}] \emph{Does leveraging both the report's title and description improve retrieval performance?}
    \item[\textbf{RQ2:}] \emph{Does the length of a report’s description affect the robustness of retrieval performance?}
    \item[\textbf{RQ3:}] \emph{How does retrieval performance change as the size of the training data increases?}
\end{itemize}

To answer these questions, we first conduct comprehensive experiments using both traditional lexical baselines and modern transformer-based sentence embedding models. Our findings demonstrate that sentence transformer models consistently outperform all other approaches. After identifying the best-performing model (\textit{multi-qa-MiniLM-L6-cos-v1}), we systematically addressed the three RQs.

First, we find that while titles alone carry meaningful semantic information, incorporating full descriptions significantly improves retrieval performance, with an average gain of 15–25 percentage points in ranking metrics. Second, our analysis shows that shorter descriptions slightly outperform longer ones, although the performance differences are minimal, suggesting that our system is robust to variations in description length. Finally, we show that retrieval performance steadily improves as the size of the training data increases. These findings highlight the feasibility and robustness of automating the linking of new reports to existing \textit{AI Incidents}, paving the way for more scalable and consistent maintenance of the AI Incident Database.

\section{Related Works}

This section reviews prior work relevant to our approach. We first examine advances in retrieval models. We then discuss how structured reporting systems have supported safety practices in fields like aviation and cybersecurity. Finally, we turn to the AI Incident Database itself, highlighting its growing importance and the need for scalable tools to support its maintenance.

\subsection{Semantic Retrieval Models}

Retrieving semantically relevant documents from large-scale textual corpora is a fundamental task in information retrieval. Traditional approaches have long relied on term-based probabilistic models such as BM25 \cite{RobertsonZ09}, which offer competitive performance and scalability. However, these methods are inherently limited by their reliance on lexical overlap, making them vulnerable to the vocabulary mismatch problem \cite{furnas1987vocabulary} and unable to capture deeper semantic relations. Earlier attempts to mitigate this limitation employed term dependency and topic modeling techniques \cite{metzler2005markov, blei2003latent, lee2000algorithms}, though these approaches proved insufficient for more nuanced semantic reasoning. More recent improvements, such as retrieval pipelines combining BM25 with T5-based document or query generation \cite{BM25T5}, seek to enrich input representations through generative query expansion.

With the advent of transformer-based language models, neural approaches to retrieval have gained prominence, enabling more expressive semantic matching through learned text representations. These methods are generally classified into sparse and dense retrieval paradigms \cite{guo2022semantic}. Sparse retrieval models improve upon classical term-based methods by enhancing token-level representations while maintaining a sparse indexing structure (e.g., DeepCT \cite{dai2020context}, docT5query \cite{nogueira2019document}). Dense retrieval methods, on the other hand, map both queries and documents into a continuous embedding space using dual-encoder architectures, computing similarity scores via a predefined function $f$. Depending on the granularity of representation, dense models can be further divided into term-level approaches — such as DC-BERT \cite{nie2020dc} and ColBERT \cite{khattab2020colbert} — which preserve fine-grained token interactions, and document-level approaches — including Sentence-BERT \cite{reimers2019sentence}, DPR \cite{karpukhin2020dense}, and the MultiQA family \cite{talmor2019multiqa} — which yield global embeddings for entire texts.

In this work, we evaluate a diverse set of retrievers in the context of the AI Incident Database. Our objective is to understand which models are most effective for linking newly submitted reports to semantically related historical \textit{AI Incidents}, where lexical overlap is often minimal and semantic cues are subtle yet crucial for accurate matching.

\subsection{Incident Databases in Safety-Critical Domains}

Several safety-critical industries have long-established incident reporting systems that serve as foundational infrastructure for risk mitigation and continuous improvement. For instance, in the aviation sector, databases maintained by agencies such as the Federal Aviation Administration (FAA)\footnote{https://www.faa.gov/data\_research/accident\_incident} and the National Aeronautics and Space Administration (NASA)\footnote{https://asrs.arc.nasa.gov/} have played a pivotal role in advancing operational safety standards \cite{RUSKIN2021100502}. These repositories systematically document accidents---defined as events resulting in serious damage or casualties---and incidents, which are precursors or near-misses that expose vulnerabilities within existing processes. Through the aggregation of structured event data, technical reports, and expert analyses, these systems have supported the development of robust safety protocols and informed regulatory decision-making.

A similar model exists in the domain of cybersecurity, where the Common Vulnerabilities and Exposures\footnote{https://cve.mitre.org/} (CVE) system provides a standardized nomenclature for publicly disclosed software vulnerabilities. Maintained by The MITRE Corporation, the CVE framework enables consistent identification, cataloging, and dissemination of security issues across heterogeneous environments.

In safety-critical applications where AI failures can lead to significant harm, the AI Incident Database\footnote{\url{https://incidentdatabase.ai/}} has recently emerged as a key resource. By systematically cataloging real-world incidents involving AI across diverse sectors, the AIID enables the identification of recurring failure modes and risk factors—critical insights for developing robust, evidence-based mitigation strategies.

\subsection{The AI Incident Database}

The AI Incident Database is a structured repository of documented failures, harms, and unintended consequences arising from the deployment of AI systems \cite{McGregor_2021}. It aims to support transparency, accountability, and risk analysis in AI development by cataloging real-world incidents reported across media sources, academic literature, and governmental investigations. Each record — referred to as an \textit{AI Incident} — denotes a case in which an AI system is implicated in causing, or nearly causing, harm. Incidents are described with rich metadata, including the organizations involved, system functionality, affected stakeholders, types of harm, and may be linked to multiple reports over time. To capture recurring patterns of failure, the AIID introduces the concept of an \textit{AI Incident Variant}, defined as an event that shares similar causes, harms, and system-level characteristics with a previously reported \textit{AI Incident}. Alongside these grounded categories, the AIID also tracks \textit{AI Issues} — speculative or potential harms that have not yet occurred or been empirically observed but are flagged as emerging concerns in the deployment of AI technologies \cite{mcgregor2022indexingairisksincidents}.

The AIID has been increasingly adopted as a foundational resource in diverse areas of AI research and governance, providing real-world evidence of failures and harms caused by deployed AI systems. Several recent works have leveraged AIID records to inform both normative and empirical investigations. For instance, \cite{rodrigues2023artificial} analyze the role of incident documentation in shaping governance responses to AI failures. Beyond analytical use, the AIID has also been integrated into regulatory and educational practices. \cite{lupo2023risky} examined its role in shaping emerging AI regulation frameworks, while \cite{feffer2023ai} evaluated its effectiveness as a pedagogical tool to build awareness of AI harms in classroom settings. The database has also been used to inform ethical risk assessments in sectors such as healthcare \cite{bondi2023taking}, cybersecurity \cite{apruzzese2023real}, and automated hiring \cite{schloetzer2023algorithmic}, emphasizing its practical relevance across domains.

Collectively, these works underscore the AIID’s growing relevance, highlighting its potential to support systematic understanding and oversight of AI harms at scale. Despite this momentum, the AIID rely on manual curation. As the volume of new reports continues to grow, there is a pressing need for scalable, automated approaches that can assist in maintaining and enriching the AIID’s structure.

In this work, we address this gap by introducing a novel retrieval-based task: automatically linking newly submitted incident reports to semantically related, pre-existing \textit{AI Incidents}. We propose a retrieval methodology designed to assist AIID maintainers in curating the database more efficiently, enabling scalable and context-aware integration of new information.

\section{Materials and Methods}

This section first introduces the dataset used in our study and the evaluation metrics employed to assess retrieval performance. We then formalize the task of associating new reports with relevant \textit{AI Incidents} as a semantic ranking problem and present the methodology adopted, including the design of the retrieval pipeline and its main processing stages.

\subsection{Dataset \& Metrics}

We conducted our experiments using data from the AI Incident Database \cite{McGregor_2021}, a publicly available repository documenting real-world failures involving AI systems. The dataset comprises user-submitted, editor-approved reports that are linked to \textit{AI Incidents}. Each report includes a title and a textual description of the event. Reports are linked to at least one corresponding incident, while a single incident may be associated with multiple reports. An illustrative example\footnote{This example is drawn directly from the AI Incident Database, available at \url{https://incidentdatabase.ai/cite/2/}.} of the relationship between an \textit{AI Incident} and two of its associated reports is presented in Table \ref{Incident-Variant Example}.

\begin{table}[h]
\begin{tabular}{p{2.5cm} p{5cm}}
    \toprule
    \parbox[t]{3cm}{\textbf{AI Incident} \\ \textbf{Title}} & Warehouse robot ruptures can of bear spray and injures workers \\ 
    \midrule
    \parbox[t]{3cm}{\textbf{AI Incident} \\ \textbf{Description}} & Twenty-four Amazon workers in New Jersey were hospitalized after a robot punctured a can of bear repellent spray in a warehouse. \\
    \midrule
    \parbox[t]{3cm}{\textbf{Report} \\ \textbf{Title \#1}} & 1 critical, 54 Amazon workers treated after bear repellent discharge in N.J. Warehouse \\
    \midrule
    \parbox[t]{3cm}{\textbf{Report} \\ \textbf{Description \#1}} & A worker at an Amazon warehouse in New Jersey was in critical condition and another 54 required treatment after being exposed to bear repellent that discharged when a can was punctured by an automated machine Wednesday morning inside the building, officials said. A total of 54 workers [...] The warehouse was cleared for re-entry around 1 p.m. by the West Windsor Health Department, but an official will revisit the building before Thursday morning as a precaution.\\
    \midrule
    \parbox[t]{3cm}{\textbf{Report} \\ \textbf{Title \#2}} & Amazon bear repellent accident sends 24 workers to the hospital \\
    \midrule
    \parbox[t]{3cm}{\textbf{Report} \\ \textbf{Description \#2}} & On Wednesday this week 24 Amazon warehouse workers in Robbinsville New Jersey were hospitalized after a robot punctured a can of repellent according to local news reports. One employee is said to be in critical condition. The accident In all, 54 workers [...] However, Amazon hardly is at fault in this instance.An Amazon spokesperson claimed: “While any serious incident is one too many, we learn and improve our programs working to prevent future incidents.” The company claims it surveys it workers every month to gauge their perception of safety in their workplace. More about bear repellent, Robots, Amazon More news from bear repellent Robots Amazon.\\
    \bottomrule
\end{tabular}
\caption{\textbf{Example of an \textit{AI Incident} and two of its corresponding reports.} The incident describes a harmful failure involving an automated system in a warehouse setting. The two reports provide independent accounts of the same underlying event, each contributing complementary details that help contextualize and document the incident.}
\label{Incident-Variant Example}
\end{table}

The dataset\footnote{We use the official AI Incident Database snapshot available at https://incidentdatabase.ai/research/snapshots/. Specifically, we rely on the backup released on October 28, 2024.} comprises 815 \textit{AI Incidents} and 3,805 reports. These records are obtained by filtering out entries labeled as \textit{AI Issues}, which — by the AIID’s definition — refer to alleged harms by an AI system that have yet to occur or be detected \cite{mcgregor2022indexingairisksincidents}. As such, they are prospective or hypothetical in nature and cannot be reliably associated with any past incident.

The performance of the proposed models is evaluated using standard ranking metrics commonly adopted in retrieval tasks: \textit{Accuracy@K}, \textit{Mean Reciprocal Rank (MRR@K)}, and \textit{Normalized Discounted Cumulative Gain (NDCG@K)}, where $K$ denotes the number of top-ranked incident predictions for a given report considered in the evaluation.

\subsection{Methodology}

In this study, we conceptualize the problem of associating newly submitted reports with existing \textit{AI Incidents} as a retrieval task. The objective is to rank \textit{AI Incidents} based on their semantic relevance to a given report.

\begin{figure*}
    \centering
    \includegraphics[width=1\linewidth]{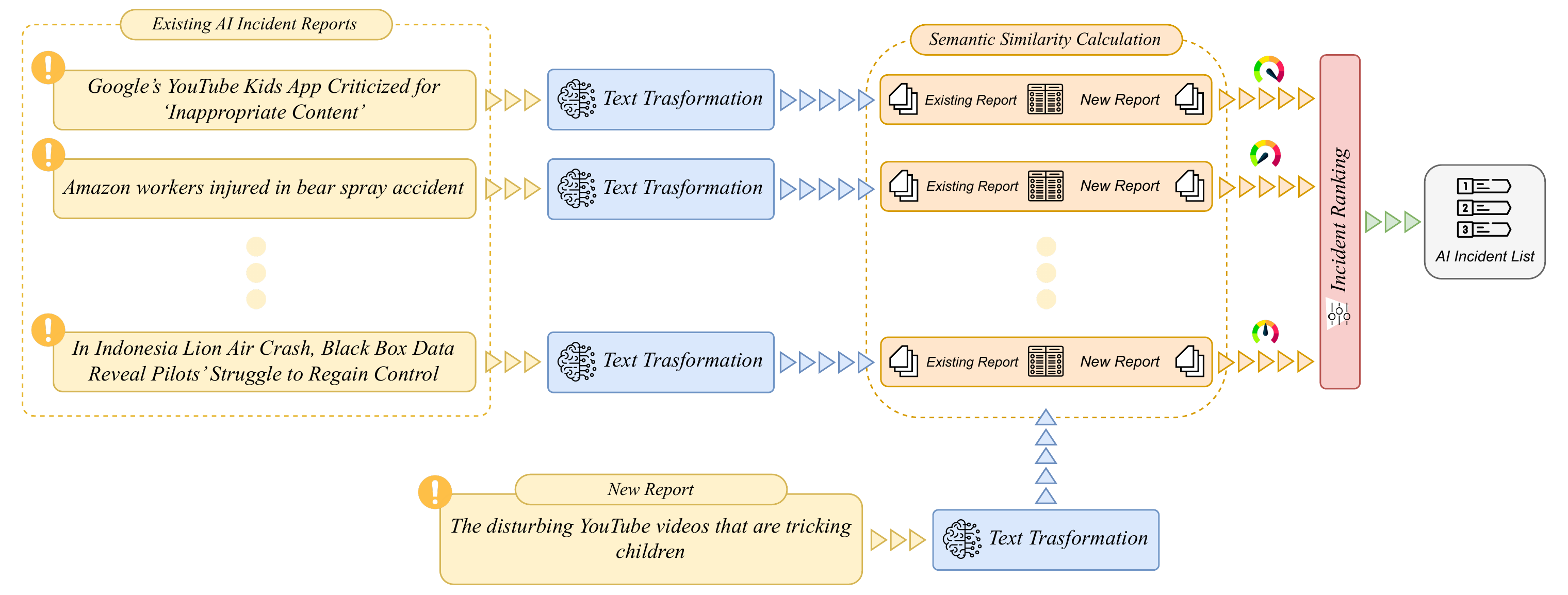}
    \caption{\textbf{Overview of the semantic retrieval pipeline for linking new reports to existing \textit{AI Incidents}.} Given a newly submitted report, the system ranks existing \textit{AI Incidents} by computing semantic similarity between their textual representations. The pipeline consists of three main stages: \textit{(1) Text Transformation}, where titles and descriptions are preprocessed and embedded; \textit{(2) Semantic Similarity Calculation}, which computes pairwise similarity scores between the report and each incident; and \textit{(3) Incident Ranking}, which generates a prioritized list of the most semantically relevant incidents.}
    \label{Pipeline}
\end{figure*}

The retrieval process follows a sequential pipeline, as detailed in Figure \ref{Pipeline}, which includes the following steps:

\begin{enumerate}
    \item \textbf{Text Transformation:} For each \textit{AI Incident} and a given report, the title and description are concatenated to form a unified textual representation for that individual entry. Subsequently, a preliminary pre-processing phase is applied, which involves standard text cleaning operations — such as the removal of non-informative characters (e.g., line breaks, emojis) — and the elimination of stopwords. The resulting texts are then converted into vector representations, according to the semantic embedding model adopted;
    \item \textbf{Semantic Similarity Calculation:} The vector representation of the considered report is systematically compared against the vector representations of all existing \textit{AI Incidents}. For each pairwise comparison, a similarity score is computed to quantify their semantic proximity;
    \item \textbf{Incident Ranking:} Based on the computed similarity scores, \textit{AI Incidents} are ranked by semantic relevance to the given report.
\end{enumerate}

The final output of the pipeline is an \textit{AI Incident List}, where existing incidents are ranked by their semantic relevance to the newly submitted report. This prioritized list is designed to assist curators in efficiently identifying the most likely matches, streamlining the process of linking new reports to documented incidents.

\section{Experiments}

In this section, we empirically evaluate the proposed retrieval framework for linking new reports to relevant \textit{AI Incidents}. We begin by outlining the experimental setup and model configurations used in our study. We then present a comparative analysis of several retrieval models to identify the most effective approach for the task. Finally, we address the three research questions, providing a detailed analysis of the factors that influence retrieval performance.

\subsection{Experimental Setup}

To assess the performance of different semantic retrieval strategies for matching \textit{AI Incidents} with the considered report, we conducted experiments using both lexical and embedding-based models. Specifically, the evaluated models include BM25 \cite{RobertsonZ09}, BM25+T5 \cite{nogueira2020document}, cross-encoders (i.e., \textit{quora-roberta-base}, \textit{ms-marco-MiniLM-L4-v2}) \cite{Lu25,Rosa22}, and sentence transformers (i.e., \textit{multi-qa-distilbert-cos-v1}, \textit{all-mpnet-base-v2}, \textit{multi-qa-MiniLM-L6-cos-v1}) \cite{ReimersG20,Song0QLL20,WangW0B0020}. Cross-encoders and sentence transformers were fine-tuned on a training set derived by splitting the dataset into 75\% for training, 12.5\% for validation, and 12.5\% for testing. All results reported in the following sections are based on the test set, which contains 475 variants. All experiments\footnote{The code will be made available upon acceptance.} were performed on a workstation equipped with an AMD Ryzen 7 5800H CPU (3.20 GHz), 32 GB RAM, and an NVIDIA RTX 3060 GPU.

\subsection{Model Selection}

Table \ref{Baselines} summarizes the performance of various models evaluated at cutoff values of  \( K = 3 \), \( K = 5 \), and \( K = 10 \). The results reveal significant distinctions among traditional lexical models, hybrid approaches, and neural embedding-based models.

Among the traditional methods, BM25 consistently exhibits the weakest performance, struggling to capture the complex semantic relationships inherent in incident-report pairs. The inclusion of synthetic queries via T5 leads to improvements in ranking accuracy, highlighting the benefits of query expansion for enhancing traditional approaches. In contrast, cross-encoder models perform suboptimally compared to the BM25+T5 approach, failing to achieve competitive results. On the other hand, sentence transformer models consistently outperform all other approaches, demonstrating superior results across all evaluation metrics. This strong performance makes them the preferred choice for addressing the research questions in this study. Specifically, we adopt the \textit{multi-qa-MiniLM-L6-cos-v1} model for subsequent analysis, as it consistently achieved the best performance across all values of $K$ and evaluation metrics considered.

\begin{table}[t]
    \scriptsize
    \begin{tabular}{@{}c@{\hspace{0.3cm}}c@{\hspace{0.3cm}}c@{\hspace{0.3cm}}c@{\hspace{0.3cm}}c@{}}  
        \toprule
        \textbf{K} & \textbf{Model} & \textbf{Accuracy@K} & \textbf{MRR@K} & \textbf{NDCG@K} \\ 
        \midrule
        \multirow{7}{*}{3} 
            & BM25 & 0.584 ± 0.022 & 0.526 ± 0.019 & 0.526 ± 0.019 \\ [1.2mm]
            & BM25+T5 & 0.704 ± 0.022 & 0.637 ± 0.017 & 0.654 ± 0.017 \\ [1.2mm]
            & quora-roberta-base & 0.513 ± 0.017 & 0.400 ± 0.012 & 0.429 ± 0.012 \\ [1.2mm]
            & ms-marco-MiniLM-L4-v2 & 0.666 ± 0.010 & 0.554 ± 0.009 & 0.583 ± 0.007 \\ [1.2mm]
            & multi-qa-distilbert-cos-v1 & 0.974 ± 0.018 & 0.948 ± 0.025 & 0.954 ± 0.023 \\ [1.2mm]
            & all-mpnet-base-v2 & 0.980 ± 0.004 & 0.955 ± 0.005 & 0.962 ± 0.004 \\ [1.2mm]
            & multi-qa-MiniLM-L6-cos-v1 & \textbf{0.982 ± 0.006} & \textbf{0.963 ± 0.010} & \textbf{0.968 ± 0.008} \\
        \midrule
        \multirow{7}{*}{5} 
            & BM25 & 0.623 ± 0.019 & 0.535 ± 0.018 & 0.557 ± 0.018 \\ [1.2mm]
            & BM25+T5 & 0.745 ± 0.019 & 0.646 ± 0.016 & 0.671 ± 0.016 \\ [1.2mm]
            & quora-roberta-base & 0.617 ± 0.018 & 0.424 ± 0.011 & 0.472 ± 0.012 \\ [1.2mm]
            & ms-marco-MiniLM-L4-v2 & 0.744 ± 0.007 & 0.572 ± 0.009 & 0.614 ± 0.008 \\ [1.2mm]
            & multi-qa-distilbert-cos-v1 & 0.982 ± 0.014 & 0.950 ± 0.023 & 0.958 ± 0.021 \\ [1.2mm]
            & all-mpnet-base-v2 & 0.985 ± 0.004 & 0.957 ± 0.004 & 0.965 ± 0.003 \\ [1.2mm]
            & multi-qa-MiniLM-L6-cos-v1 & \textbf{0.987 ± 0.005} & \textbf{0.965 ± 0.010} & \textbf{0.970 ± 0.008}\\
        \midrule
        \multirow{7}{*}{10} 
            & BM25 & 0.666 ± 0.017 & 0.541 ± 0.017 & 0.571 ± 0.017 \\ [1.2mm]
            & BM25+T5 & 0.790 ± 0.014 & 0.653 ± 0.016 & 0.686 ± 0.015 \\ [1.2mm]
            & quora-roberta-base & 0.741 ± 0.010 & 0.440 ± 0.011 & 0.505 ± 0.010 \\ [1.2mm]
            & ms-marco-MiniLM-L4-v2 & 0.837 ± 0.010 & 0.584 ± 0.009 & 0.645 ± 0.008 \\ [1.2mm]
            & multi-qa-distilbert-cos-v1 & 0.988 ± 0.009 & 0.950 ± 0.023 & 0.960 ± 0.019 \\ [1.2mm]
            & all-mpnet-base-v2 & 0.989 ± 0.002 & 0.957 ± 0.004 & 0.965 ± 0.003 \\ [1.2mm]
            & multi-qa-MiniLM-L6-cos-v1 & \textbf{0.990 ± 0.005} & \textbf{0.965 ± 0.010} & \textbf{0.971 ± 0.008} \\
        \bottomrule
    \end{tabular}   
    \caption{\textbf{Performance comparison of baseline and embedding-based models across different values of K.} \textit{multi-qa-MiniLM-L6-cos-v1} model consistently achieved the best performance across all values of $K$ and evaluation metrics.}
    \label{Baselines}
\end{table}

\subsection{Impact of Title and Description Combination on Retrieval Performance (RQ1)}

In the \textit{AIID} system, reports are described by a title and an associated textual description. However, user-provided descriptions are only required to exceed a minimum length of 80 characters, a constraint that does not guarantee sufficient detail or informativeness for retrieval tasks. To address this limitation, we investigated the extent to which combining the title and description of each report enhances top-K retrieval performance, particularly in scenarios where individual descriptions may lack sufficient detail.

As shown in Table \ref{RQ1}, incorporating both the title and description consistently yields superior retrieval performance across all considered values of \(K\) and evaluation metrics. For instance, in terms of Accuracy@K, using only the title achieves a score of 0.772 ± 0.012 at \(K=3\), while combining title and description raises the performance to 0.982 ± 0.006. A similar trend is observed at \(K=5\), where the title-only configuration attains an Accuracy@K of 0.808 ± 0.015, compared to 0.987 ± 0.005 when both components are utilized.

Overall, these results emphasize that leveraging both title and description results in an average improvement across all evaluation criteria. Specifically, a 15\% gain was observed in Accuracy@K with \(K=10\), while a peak improvement of 25\% was observed in MRR@K with \(K=3\).

\begin{table}[t]
    \centering
    \scriptsize
    \begin{tabular}{@{}c@{\hspace{0.3cm}}c@{\hspace{0.3cm}}c@{\hspace{0.3cm}}c@{\hspace{0.3cm}}c@{}}
        \toprule
        \textbf{K} & \textbf{Input Type} & \textbf{Accuracy@K} & \textbf{MRR@K} & \textbf{NDCG@K} \\ 
        \midrule
        \multirow{2}{*}{3} 
            & Incident Title & 0.772 ± 0.012 & 0.707 ± 0.013 & 0.724 ± 0.013 \\ [1.2mm]
            & Incident Title + Description & \textbf{0.982 ± 0.006} & \textbf{0.963 ± 0.010} & \textbf{0.968 ± 0.008} \\
        \midrule
        \multirow{2}{*}{5} 
            & Incident Title & 0.808 ± 0.015 & 0.715 ± 0.013 & 0.739 ± 0.014 \\ [1.2mm]
            & Incident Title + Description & \textbf{0.987 ± 0.005} & \textbf{0.965 ± 0.010} & \textbf{0.970 ± 0.008} \\
        \midrule
        \multirow{2}{*}{10} 
            & Incident Title  & 0.855 ± 0.021 & 0.722 ± 0.014 & 0.754 ± 0.015 \\ [1.2mm]
            & Incident Title  + Description & \textbf{0.990 ± 0.005} & \textbf{0.965 ± 0.010} & \textbf{0.971 ± 0.008} \\
        \bottomrule
    \end{tabular}
    \caption{\textbf{Retrieval performance comparison using title only versus title combined with description.} Combining title and description consistently improves retrieval effectiveness across all evaluation metrics and values of K, with gains ranging from 15\% to 25\% compared to using the title alone.}
    \label{RQ1}
\end{table}

\subsection{Impact of Description Length on Retrieval Robustness (RQ2)}

In the \textit{AIID} system, reports are associated with textual descriptions, whose lengths can vary considerably. This variability introduces potential challenges, as excessively short descriptions may lack sufficient detail, while overly long descriptions may contain noise that negatively impacts retrieval performance. To address this, we examine if the length of descriptions influences performance, assessing whether shorter or longer descriptions produce different results for our task.

Reports were stratified into subsets according to two distinct length-based partitioning criteria: one based on the median description length and the other on the 25th percentile. A report is classified as having a short description if its length falls below the respective threshold (median or 25th percentile), and as having a long description otherwise.

As shown in Table \ref{RQ2}, shorter descriptions yield slightly better retrieval performance across all evaluation metrics. For instance, fixing \( K=3 \), reports with descriptions shorter than the median achieve an MRR@K of 0.979 ± 0.010, compared to 0.949 ± 0.013 for longer descriptions. A similar trend is observed when using the 25th percentile threshold: reports below this threshold reach an MRR@K of 0.978 ± 0.014, while those above attain 0.959 ± 0.013, respectively.

Although slight improvements are observed, the performance differences between shorter and longer descriptions remain minimal, suggesting that both types provide comparable retrieval effectiveness. This finding indicates that our framework maintains robust performance regardless of description length, demonstrating resilience to variability in the level of detail provided in the reports.

\begin{table}[t]
    \centering
    \resizebox{\linewidth}{!}{%
    \begin{tabular}{@{}c@{\hspace{0.6cm}}c@{\hspace{0.6cm}}c@{\hspace{0.6cm}}c@{\hspace{0.6cm}}c@{}}
        \toprule
        \textbf{K} & \textbf{Length Condition} & \textbf{Accuracy@K} & \textbf{MRR@K} & \textbf{NDCG@K} \\ 
        \midrule
        \multirow{5}{*}{3} 
            & < Median & \textbf{0.989 ± 0.004} & \textbf{0.979 ± 0.010}& \textbf{0.965 ± 0.003} \\ [1.2mm]
            & $\geq$ Median & 0.973 ± 0.009 & 0.949 ± 0.013 & 0.955 ± 0.012 \\
        \cmidrule(lr){2-5}
            & < 25th Percentile & \textbf{0.988 ± 0.008} & \textbf{0.978 ± 0.014} & \textbf{0.981 ± 0.012} \\ [1.2mm]
            & $\geq$ 25th Percentile & 0.979 ± 0.006 & 0.959 ± 0.010 & 0.964 ± 0.008 \\
        \midrule
        \multirow{5}{*}{5} 
            & < Median & \textbf{0.992 ± 0.005} & \textbf{0.980 ± 0.010} & \textbf{0.983 ± 0.009} \\ [1.2mm]
            & $\geq$ Median & 0.982 ± 0.008 & 0.951 ± 0.013 & 0.959 ± 0.011 \\
        \cmidrule(lr){2-5}
            & < 25th Percentile & \textbf{0.988 ± 0.008} & \textbf{0.978 ± 0.014} & \textbf{0.981 ± 0.012} \\ [1.2mm]
            & $\geq$ 25th Percentile & 0.987 ± 0.005 & 0.961 ± 0.010 & 0.967 ± 0.008 \\
        \midrule
        \multirow{5}{*}{10} 
            & < Median & \textbf{0.992 ± 0.005} & \textbf{0.980 ± 0.010} & \textbf{0.983 ± 0.009} \\ [1.2mm]
            & $\geq$ Median & 0.987 ± 0.009 & 0.952 ± 0.013 & 0.960 ± 0.010 \\
        \cmidrule(lr){2-5}
            & < 25th Percentile & 0.988 ± 0.008 & \textbf{0.978 ± 0.014} & \textbf{0.981 ± 0.012} \\ [1.2mm]
            & $\geq$ 25th Percentile & \textbf{0.990 ± 0.006} & 0.962 ± 0.010 & 0.969 ± 0.008 \\
        \bottomrule
    \end{tabular}%
    }
    \caption{\textbf{Retrieval performance across different description lengths.} Descriptions are classified as shorter or longer based on whether their length falls below or above the median or 25th percentile thresholds. Performance differences across these groups are minimal, suggesting that the retrieval framework is robust to input length variability.}
    \label{RQ2}
\end{table}

\subsection{Impact of Training Data Size on Retrieval Performance (RQ3)}

\begin{figure*}
    \centering
    \includegraphics[width=1\linewidth]{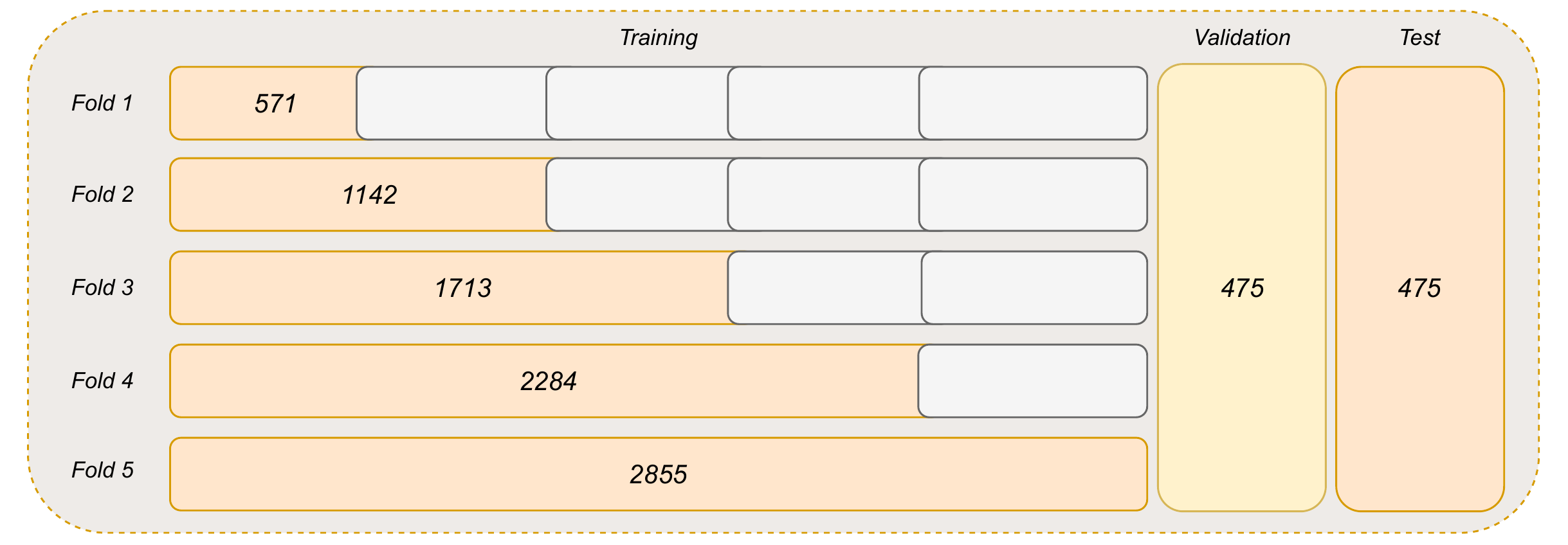}
    \caption{\textbf{Evaluation protocol for assessing the impact of training data scale.} The training set is chronologically ordered and incrementally expanded in five cumulative folds, each adding a new batch of 571 newly reported reports. The validation and test sets remain fixed across all folds, enabling consistent evaluation of how retrieval performance changes as more data becomes available.}
    \label{Fold}
\end{figure*}

Until this point, our evaluation relied on static validation strategies with a fixed training set. However, considering that new reports are continuously reported over time, it becomes essential to assess how the model behaves as additional training data becomes available. To this end, we designed a progressive training protocol with fixed validation and test sets. Figure \ref{Fold} illustrates the overall structure of the protocol. Specifically, the training set was first temporally ordered according to the reporting time of each instance. The ordered training set was then partitioned into five folds. Each fold $F_i$ (for $i = 1, \dots, 5$) includes both the instances from fold $F_{i-1}$ and an additional batch of newly introduced instances. Formally, if each batch contains $N$ new instances, the size of the $i$-th temporal fold is:

\[
|F_i| = i \times N
\]

with $F_1$ consisting of $N$ instances, $F_2$ containing $2N$ instances (the first $N$ plus an additional $N$), and so on. In our setup, each batch consists of 571 instances, resulting in fold sizes of 571, 1142, 1713, 2284, and 2855 instances, respectively.

We fine-tuned a separate model on each fold to evaluate how retrieval performance evolves as the size of the training set increases.

As shown in Table \ref{RQ3}, the model demonstrates a consistent performance improvement as additional training data is incorporated, suggesting a beneficial effect from exposure to a growing number of incidents. For $K = 3$, accuracy increases from 0.921 in the first fold to 0.953 in the final one. Corresponding gains are observed in MRR, which rises from 0.859 to 0.926, and in NDCG, from 0.875 to 0.933. These trends indicate a progressively enhanced ranking capability as the training set expands over time. A similar pattern emerges for $K = 5$, where accuracy improves from 0.953 to 0.962 across folds. Similarly, MRR increases from 0.866 to 0.928, while NDCG advances from 0.888 to 0.937. Overall, these findings highlight the scalability and stability of our system under realistic data growth conditions.

\begin{table}[t]
    \centering
    \begin{tabular}{@{}c@{\hspace{0.6cm}}c@{\hspace{0.6cm}}c@{\hspace{0.6cm}}c@{\hspace{0.6cm}}c@{}}
        \toprule
        \textbf{K} & \textbf{Fold} & \textbf{Accuracy@K} & \textbf{MRR@K} & \textbf{NDCG@K} \\ 
        \midrule
        \multirow{6}{*}{3} 
        & 1 & 0.921 & 0.859 & 0.875 \\ [1.2mm]
        & 2 & 0.940 & 0.878 & 0.894 \\ [1.2mm]
        & 3 & 0.883 & 0.808 & 0.827 \\ [1.2mm]
        & 4 & 0.953 & 0.925 & 0.932 \\ [1.2mm]
        & 5 & \textbf{0.953 (+3.47\%)}  & \textbf{0.926 (+7.80\%)} & \textbf{0.933 (+6.63\%) } \\
        \midrule
        \multirow{5}{*}{5} 
        & 1 & 0.953 & 0.866 & 0.888 \\ [1.2mm]
        & 2 & 0.968 & 0.885 & 0.906 \\ [1.2mm]
        & 3 & 0.899 & 0.812 & 0.834 \\ [1.2mm]
        & 4 & 0.972 & 0.929 & 0.940 \\ [1.2mm]
        & 5 & \textbf{0.962 (+0.95\%)}  & \textbf{0.928 (+7.16\%)} & \textbf{0.937 (+5.52\%)} \\
        \midrule
        \multirow{5}{*}{10} 
        & 1 & 0.975 & 0.870 & 0.895 \\ [1.2mm]
        & 2 & 0.990 & 0.888 & 0.913 \\ [1.2mm]
        & 3 & 0.931 & 0.816 & 0.844 \\ [1.2mm]
        & 4 & 0.987 & 0.931 & 0.944 \\ [1.2mm]
        & 5 & 0.975 (+0.00\%)   & \textbf{0.930 (+6.90\%)}  & \textbf{0.941 (+5.14\%)}  \\
        \bottomrule
    \end{tabular}
    \caption{\textbf{Retrieval performance across temporally expanding training folds.} Each fold incrementally introduces new training data in chronological order. Retrieval performance improves consistently across folds and metrics, highlighting the model’s stability to newly emerging incident reports.}
    \label{RQ3}
\end{table}

\section{Conclusions, Limitations and Future Works}

This work addresses the critical challenge of automating the association between newly reported reports and previously documented \textit{AI Incidents} in the AI Incident Database. This task, currently performed manually, poses significant scalability limitations, potentially introducing also inconsistencies. To overcome these issues, we formalize the report-to-incident association as a semantic retrieval task and propose a framework that ranks historical incidents based on their semantic similarity to a given report.

We evaluate our approach using the AIID dataset, comprising 815 incidents and 3,805 reports. After benchmarking a range of retrieval models, we identify the \textit{multi-qa-MiniLM-L6-cos-v1} sentence transformer as the most effective, consistently outperforming all baselines across Accuracy@K, MRR@K, and NDCG@K.

Our experimental analysis yields three key findings. First, combining the title and description of reports significantly improves retrieval performance over using the title alone, with gains ranging from 15\% to 25\% (RQ1). Second, the model maintains robust performance regardless of description length, with only marginal differences observed between shorter and longer inputs—highlighting the resilience of the framework to variability in input verbosity (RQ2). Third, our analysis shows that retrieval performance consistently improves as more training data becomes available, with gains observed across all evaluation metrics. This indicates that the proposed system scales effectively and remains robust as the incident repository grows over time (RQ3).

Despite these promising results, our framework relies exclusively on the textual content of reports, disregarding potentially valuable metadata such as incident categories, tags, or sources. This constraint may limit the model's ability to capture deeper contextual or structural cues. Furthermore, by focusing solely on semantic similarity, the approach may overlook latent factors such as causal mechanisms, domain-specific nuances, or system-level attributes that are not explicitly described in the report text. Moreover, the current evaluation is restricted to pairwise comparisons between a single report and individual candidate incidents, without considering the relational structure that may emerge from groups of reports referring to the same underlying incident.

Future work will explore the integration of structured metadata to complement text-based similarity, as well as methods for explainable retrieval to enhance transparency in the report-to-incident mapping. Another promising direction would be the integration of large language models (LLMs), leveraging their advanced reasoning and summarization capabilities to identify deeper semantic and causal links between reports that go beyond surface-level textual similarity. Finally, it would be valuable to evaluate our approach in the context of the actual AIID workflow, assessing its effectiveness in supporting editors during the real-time curation of incident reports.

\bibliography{mybibfile}

\end{document}